\documentclass[12pt]{article}

\usepackage{graphicx}
\graphicspath{%
   {converted_graphics/}
    {uv-plane2.jpg}
}

\begin{document}

\title{Classical and Quantum Properties of a Two-Sphere Singularity} 





\author{T.M. Helliwell \\
              Department of Physics, Harvey Mudd College, \\ Claremont, CA 91711 USA 
           \and
           D.A. Konkowski \\
              Department of Mathematics, U.S. Naval Academy, \\ Annapolis, MD 21402 USA}


\maketitle

\begin{abstract}

\noindent Recently B\"ohmer and Lobo have shown that a metric due to Florides, which has been used as an interior Schwarzschild solution, can be extended to reveal a classical singularity that has the form of a two-sphere. Here the singularity is shown to be a naked scalar curvature singularity that is both timelike and gravitationally weak. It is also shown to be a quantum singularity because the Klein-Gordon operator associated with quantum mechanical particles approaching the singularity is not essentially self-adjoint.

\end{abstract}


\noindent 

\section{Introduction}    

An unusual singularity has been described by B\"ohmer and Lobo \cite{BL}. They studied a constant-density version of a spherically-symmetric spacetime due to Florides \cite{F}, which has been used as an interior Schwarzschild solution with vanishing radial pressure, and which can be interpreted as an ``Einstein cluster" \cite{E}.  It has the metric

\begin{equation}
ds^2 = -\frac{dt^2}{\sqrt{1 - (r/R)^2}} + \frac{dr^2}{1 - (r/R)^2} + r^2 d\Omega^2,
\end{equation}

\noindent where $R = \sqrt{3/8 \pi \rho_0}$ in terms of the constant energy density $\rho_0$, and $d\Omega^2 = d\theta^2 + \sin^2 \theta d\phi^2$. The coordinate ranges are $r < R$, $- \infty < t < \infty $, $0 \le \theta \le \pi$, and $0 \le \phi < 2\pi$. 

B\"ohmer and Lobo transform to a new coordinate $\alpha = \sin^{-1}(r/R)$ and then rescale $t$ to give

\begin{equation}
ds^2 = -\frac{dt^2}{\cos\alpha} + R^2 d\alpha^2 + R^2 \sin^2\alpha \  d\Omega^2.
\end{equation}

\noindent The spatial portion of the global extension of this solution can be considered to be a three-sphere containing a single equatorial two-sphere. (In what follows however we interpret the angle $\alpha$ to be a radial coordinate and the spatial part of the metric to be the well-known "round metric" of a three-sphere.) The radial coordinate $\alpha$ can either take the values $0 < \alpha \le \pi/2$ (half a three-sphere) or  $- \pi/2 \le \alpha \le \pi/2$ (two half three-spheres joined at $\alpha = 0$ with $\alpha = - \pi /2$ identified with $\alpha = + \pi/2.$) \footnote{See Figures 1 and 2 in B\"ohmer and Lobo.}

\par The B\"ohmer-Lobo spacetime is static, spherically symmetric, regular at $\alpha = 0$, and it has vanishing radial stresses \cite{BL}. It is also Petrov Type D and Segre Type A1 ([(11) 1, 1]) \footnote{calculated using CLASSI}, and it satisfies the strong energy condition automatically and the dominate energy condition with certain more stringent requirements \cite{F}.  Vertical cuts through the three-sphere define latitudinal two-spheres; in particular, the equatorial cut at $\alpha = \pi/2$ is a two-sphere on which scalar polynomial invariants diverge and the tangential pressure diverges as well. In the following sections we will explore the classical and quantum singularity structure of this spacetime.

\section{Classical singularities}

We use a variation of the Ellis and Schmidt classification scheme \cite{ES} to define singular points as the endpoints of incomplete geodesics in maximal spacetimes. Singularities come in many types \cite{ES}: the strongest are scalar curvature singularities, in which  approaching particles experience infinite tidal forces and there is at least one scalar quantity constructed from the metric tensor $g_{ab}$, the antisymmetric tensor $\eta_{abcd}$, and the Riemann tensor $R_{abcd}$,  along an incomplete geodesic ending at a point $q$, which is unbounded as the geodesic approaches $q$ (see, for example, \cite{ES} and \cite{HE}).

B\"ohmer and Lobo show that scalar polynomial invariants of the Riemann tensor diverge at $\alpha = \pi/2$. To verify that this is a true singularity we must also show that it can be reached by causal geodesics. The  spacetime is spherically symmetric, so it is sufficient to study geodesics in the equatorial $(\theta = \pi/2)$ plane. A complete set of first integrals of the timelike (- 1) or null (0) geodesic equations is \cite{BL}

\begin{equation}
\dot t = \epsilon \cos \alpha  \ \ \ \ \ \ \ \dot \phi = \frac{\ell}{\sin^2 \alpha}
\end{equation}

\begin{equation}
R^2 \dot \alpha^2 + \left( - \epsilon^2\cos \alpha +  \frac{R^2\ell^2}{\sin^2 \alpha}\right) =  \{- 1, 0 \}
\end{equation}
 
\noindent where $\epsilon$ and $\ell$ are constants, with $\epsilon \ge 1$. We can therefore identify an effective potential
 
 \begin{equation}
V_{\mathrm{eff}} =  - \epsilon^2\cos \alpha + \frac{R^2\ell^2}{\sin^2 \alpha} 
\end{equation}

\noindent which, in the case of radial ($\ell = 0$) geodesics, increases from $-\epsilon^2$ to zero as $\alpha$ increases from 0 to $\pi/2$. Therefore radial timelike geodesics beginning at small $\alpha$ cannot reach $\alpha = \pi/2$; they are reflected by the potential barrier back to small $\alpha$ at $\alpha = \cos^{-1}(1/\epsilon^2)$. (Geodesics with $\ell \ne 0$ are turned back at even smaller values of $\alpha$.) Therefore the apparent singularity at $\alpha = \pi/2$ is timelike geodesically complete, so timelike geodesic observers never ``fall into" the singularity at $\alpha = \pi/2$.

\par The equation of motion for radial \emph{null} geodesics is 

\begin{equation}
R^{2} \dot \alpha^2 + V_{\mathrm{eff}} = 0 
\end{equation} 

\noindent with the same effective potential, so they are able to penetrate all the way to the singularity. In fact, the affine parameter $\lambda$ along a null geodesic is finite as $\alpha \rightarrow  \pi/2$, since

\begin{equation}
\lambda = \frac{R}{\epsilon} \int_0^{\pi / 2} \frac{d\alpha}{\sqrt{\cos \alpha}} < \infty.  
\end{equation} 

\noindent The spacetime is therefore null geodesically incomplete at $\alpha = \pi/2$. This incompleteness is the necessary condition to confirm that  the two-sphere at $\alpha = \pi/2$ is singular. 

\par We would also like to know if the singularity is timelike, spacelike, or null. In double-null form the B\"ohmer-Lobo metric becomes 

\begin{equation}
ds^2 = -2e^{- 2f(u, v)}dudv + r^2d\Omega^2,
\end{equation}

\noindent where $u = t + g(r)$ and $v = t - g(r)$, with $g(r) = \int_{r_0}^r {dr (1 - (r/R)^2)^{- 1/4}}$. If $g(r)$ goes to infinity at the singularity, the singularity is null, but if $g(r)$ remains finite the singularity is timelike (see, e.g., Lake\cite{L}). Here $g(r)$ remains finite as $r \rightarrow R$, so the singularity is \emph{timelike}, as illustrated in Fig. 1. It is also clearly \emph{naked}.

\begin{figure}[tbp] 
  \centering
  \includegraphics[width=5.67in,height=3.54in,keepaspectratio]{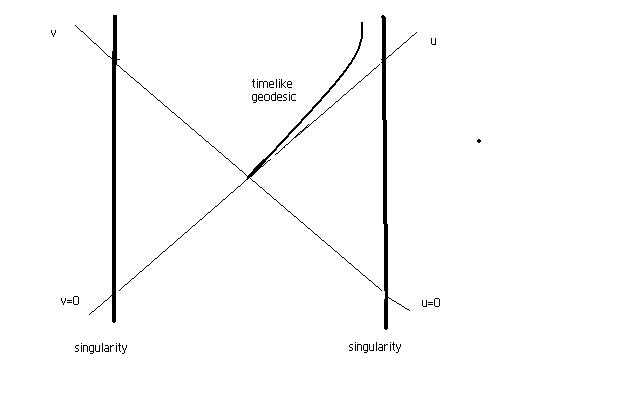}
  \caption{This is the uv-plane; it indicates the singularities in the B\"ohmer-Lobo spacetime and a representative timelike geodesic which is repelled from the singularity.}
  \label{fig:uv-plane}
\end{figure}

\par It is also interesting to know if the two-sphere is a strong or weak singularity. In the case of null geodesics a singularity is Tipler-strong if an area tangential to the geodesics, with sides represented by the tangential Jacobi fields, goes to zero as the singularity is approached \cite{ES, N1}. Nolan \cite{N} has shown that in spherically symmetric geometries the tangential Jacobi fields have the norm

\begin{equation}
\eta(\lambda) = r(\lambda)\int_{\lambda_0}^{\lambda} \frac{d\lambda'}{r^2(\lambda')}
\end{equation}

\noindent where $r$ is the radius and $\lambda$ is an affine parameter along the null geodesic. These integrals are finite and nonzero, so there is no strong singularity. The two-sphere singularity is therefore gravitationally weak, which is in accord with a general theorem of Nolan that a radial null geodesic terminating at a non-central $(r \ne 0)$ singularity terminates in a gravitationally weak singularity \cite{N1}.

\section{Quantum singularities}

Massive or massless test particles following timelike or null geodesics play an essential role in defining classical singularities. Classical particles do not exist, however, which suggests the need to find a better definition of singularities.  Horowitz and Marolf \cite{HM} have proposed the following procedure for using quantum mechanical particles to identify singularities: They define a spacetime to be quantum mechanically \emph{non}singular if the evolution of a test scalar wave packet, representing a quantum particle, is uniquely determined by the initial wave packet, the manifold, and the metric without having to place arbitrary boundary conditions at the classical singularity.  

\par If a quantum particle approaches a quantum singularity, however, its wave function may change in an indeterminate way; it may even be absorbed or another particle emitted.  This is a close analog to the definition of classical singularities: A classical singularity, as the endpoint of geodesics, can affect a classical particle in an arbitrary way; it can, for example, absorb (or not) an approaching particle, and can emit (or not) some other particle, undetermined by what comes before in spacetime. 
  
\par  Mathematically, the evolution of a quantum wave packet is related to properties of the appropriate quantum mechanical operator.  Horowitz and Marolf therefore define a static spacetime to be quantum mechanically singular \cite{HM} if the spatial portion of the Klein-Gordon operator is not essentially self-adjoint \cite{RS, Rich}. In this case the evolution of a test scalar wave packet is not determined uniquely by the initial wave packet; boundary conditions at the classical singularity are needed to ``pick out" the correct wavefunction, and thus one needs to add information that is not already present in the wave operator, spacetime metric and manifold.  Horowitz and Marolf \cite{HM} showed that although some classically singular spacetimes are quantum mechanically singular as well, others are quantum mechanically $non$singular.  A number of papers have tested additional spacetimes to see whether or not the use of quantum particles ``heals'' their classical singularities \cite{KH, HKA, KHW, MM, BFW, PL1, PL2, PL3, HK, KRHW}.  

\par   One way to test for essential self-adjointness is to use the von Neumann criterion of deficiency indices \cite{VN, weyl}, which involves studying solutions to the equation $A\Psi = \pm i\Psi$, where $A$ is the spatial Klein-Gordon operator, and finding the number of solutions that are square integrable ($i.e.$, $\in \mathcal{L}^2(\Sigma)$ on a spatial slice $\Sigma$) for each sign of $i$.  Another approach, which we have used before \cite{RS, KHW, KRHW} and will use here, has a more direct physical interpretation.  A theorem of Weyl \cite{RS, weyl} relates the essential self-adjointness of the Hamiltonian operator to the behavior of the ``potential" in an effective one-dimensional Schr\"odinger equation, which in turn determines the behavior of the scalar-wave packet.  The effect is determined by a \emph{limit point-limit circle} criterion. 

After separating the wave equation for the static, spherically-symmetric metric, with changes in both dependent and independent variables, the radial equation can be written as a one-dimensional Schr\"odinger equation $Hu(x) = Eu(x)$  where the operator $H=-d^2/dx^2 + V(x)$ and $E$ is a constant, and any singularity is assumed to be at $x = 0$.  This form allows us to use the limit point-limit circle criteria described in Reed and Simon \cite{RS}.\\  

\noindent $\mathbf{Definition}$.  \emph{The potential} $V(x)$  \emph{is in the limit circle case at}  $x = 0$ \emph{if for some, and therefore for all} $E$, \emph{all solutions of} $Hu(x) = Eu(x)$ \emph {are square integrable at zero.  If} $V(x)$ \emph{is not in the limit circle case, it is in the limit point case.}\\

\par There are of course two linearly independent solutions of the Schr\"odinger equation for given $E$. If $V(x)$ is in the limit circle case at zero, both solutions are square integrable ($\in \mathcal{L}^2(\Sigma)$)   at zero, so all linear combinations $\in \mathcal{L}^2(\Sigma)$  as well.  We would therefore need a boundary condition at  $x=0$ to establish a unique solution.  If $V(x)$ is in the limit \emph{point} case, the  $\mathcal{L}^2(\Sigma)$    requirement eliminates one of the solutions, leaving a unique solution without the need of establishing a boundary condition at $x=0$.  This is the whole idea of testing for quantum singularities; there is no singularity if the solution in unique, as it is in the limit point case.  A useful theorem is the following.\\

\noindent $\mathbf{Theorem}$ (Theorem X.10 of Reed and Simon \cite{RS}).  \emph{Let} $V(x)$ \emph{be continuous and positive near zero.  If} $V(x) \ge\frac{3}{4}x^{-2}$ \emph{near zero then} $V(x)$ \emph{is in the limit point case.  If for some} $\epsilon>0$, $V(x) \le(\frac{3}{4}-\epsilon)x^{-2}$ \emph{near zero, then} $V(x)$ \emph{is in the limit circle case.}\\

\noindent The theorem states in effect that the potential is only limit point if it is sufficiently repulsive at the origin that one of the two solutions of the one-dimensional Schr\"odinger equation blows up so quickly that it fails to be square integrable. 
 
The Klein-Gordon equation

\begin{equation}
|g|^{-1/2}\left(|g|^{1/2}g^{\mu \nu} \Phi,_{\nu}\right),_{\mu} = M^2 \Phi
\end{equation}

\noindent for a scalar function $\Phi$ has mode solutions of the form

\begin{equation}
\Phi \sim e^{- i \omega t} F(\alpha) Y_{\ell m}(\theta, \phi)
\end{equation}

\noindent for spherically symmetric metrics, where the $Y_{\ell m}$ are spherical harmonics and $\alpha$ is the radial coordinate. The radial function $F(\alpha)$ for the B\"ohmer-Lobo metric obeys

\begin{equation}
F'' + \left(2\cot\alpha + \frac{1}{2}\tan\alpha\right) F' + \left[R^2 \omega^2 \cos\alpha - \frac{\ell(\ell + 1)}{\sin^2\alpha} - R^2 M^2\right] F = 0,
\end{equation}

\noindent and square integrability is judged by finiteness of the integral

\begin{equation}
I = \int d\alpha d\theta d\phi \sqrt{\frac{g_3}{g_{00}}} \Phi^* \Phi ,
\end{equation}

\noindent where $g_3$ is the determinant of the spatial metric. The substitutions $z = \pi/2 - \alpha$, to place the singularity at $z = 0$, and $F(\alpha) = R^{-3/2} (\cos z)^{-1} \psi (x)$, where $x = \int^z dz \sqrt{\sin z}$,  convert the integral and differential equation to the one-dimensional Schr\"odinger forms $\int dx \psi^* \psi$ and 

\begin{equation}
\frac{d^2 \psi}{dx^2} + (E - V)\psi = 0,
\end{equation}

\noindent where $E = R^2 \omega^2$ and 

\begin{equation}
V = \frac{R^2 M^2}{\sin z}+ \frac{\ell(\ell + 1)}{\sin z \cos^2 z}.
\end{equation}

\noindent For small $z$, $x = \int_0^z dz \sqrt{\sin z} \sim z^{3/2}$, so the potential as $x \rightarrow 0$ is

\begin{equation}
V(x) \sim \frac{R^2 M^2 + \ell (\ell + 1)}{x^{2/3}} < \frac{3}{4x^2}.
\end{equation}

\noindent It follows from the theorem that $V(x)$ is in the limit circle case, so $x = 0$ is a quantum singularity. The Klein-Gordon operator is therefore not essentially self-adjoint. Quantum mechanics fails to heal the singularity.

Finally, the fact that the singularity is gravitationally weak suggests that an extension through the singularity might be possible \cite{N}. For example, using the two half-sphere version of the B\"ohmer-Lobo geometry, one might be able to extend the spacetime through the hypersurface that identifies $- \pi/2$ with $+\pi/2$. Null geodesics could then penetrate the hypersurface, whereas timelike geodesics are trapped in the half-sphere. We have not explored the differentiability of such an extension because the fact that the hypersurface is quantum mechanically singular means that any possible extension would be of little or no physical interest.   

\section*{Acknowledgements}
One of us (DAK) would like to thank Queen Mary, University of London for their hospitality while finishing this manuscript. We are also grateful to the referees for helpful comments.

\end{document}